\documentclass[12pt, preprint]{aastex}  

\shorttitle{CSE around Galactic Cepheids: Y~Oph and $\alpha$~Per}  
\shortauthors{M\'erand et al.}  
  
\begin{document}  
  
\author{Antoine~M\'erand}  
\affil{Center for High Angular Resolution Astronomy, Georgia State  
  University, PO Box 3965, Atlanta, Georgia 30302-3965, USA}  
\email{antoine@chara-array.org}  
    
\author{Jason~P.~Aufdenberg}  
\affil{Embry-Riddle Aeronautical  
  University, Physical Sciences Department, 600 S. Clyde Morris Blvd,  
  Daytona Beach, FL 32114, USA}  
  
\author{Pierre~Kervella and Vincent~Coud\'e~du~Foresto}  
\affil{LESIA, UMR 8109, Observatoire de Paris,  
  5 place Jules Janssen, 92195 Meudon, France}   
  
\author{ Theo~A.~ten~Brummelaar,   
  Harold~A.~McAlister,  
  Laszlo~Sturmann,  
  Judit~Sturmann and  
  Nils~H.~Turner}  
\affil{Center for High Angular Resolution Astronomy, Georgia State  
  University, PO Box 3965, Atlanta, Georgia 30302-3965, USA}  
  
\title{Extended envelopes around Galactic Cepheids III. Y~Oph and  
$\alpha$~Per from near-infrared interferometry with CHARA/FLUOR}

\begin{abstract}  
Unbiased angular diameter measurements are required for accurate
distances to Cepheids using the interferometric Baade Wesselink method
(IBWM). The precision of this technique is currently limited by
interferometric measurements at the 1.5\% level. At this level, the
center-to-limb darkening (CLD) and the presence of circumstellar
envelopes (CSE) seem to be the two main sources of bias. The
observations we performed aim at improving our knowledge of the
interferometric visibility profile of Cepheids. In particular, we
assess the systematic presence of CSE around Cepheids in order
determine accurate distances with the IBWM free from CSE biased
angular diameters.  We observed a Cepheid (Y~Oph) for which the
pulsation is well resolved and a non-pulsating yellow supergiant
($\alpha$~Per) using long-baseline near-infrared interferometry. We
interpreted these data using a simple CSE model we previously
developed. We found that our observations of $\alpha$~Per do not
provide evidence for a CSE. The measured CLD is explained by an
hydrostatic photospheric model. Our observations of Y~Oph, when
compared to smaller baseline measurements, suggest that it is
surrounded by a CSE with similar characteristics to CSE found
previously around other Cepheids. We have determined the distance to
Y~Oph to be $d=491\pm18\;\mathrm{pc}$. Additional evidence points
toward the conclusion that most Cepheids are surrounded by faint CSE,
detected by near infrared interferometry: after observing four
Cepheids, all show evidence for a CSE. Our CSE non-detection around a
non-pulsating supergiant in the instability strip, $\alpha$~Per,
provides confidence in the detection technique and suggests a
pulsation driven mass-loss mechanism for the Cepheids. 
\end{abstract}  
      
  
\maketitle  
  
\section{Introduction}  
  
In our two previous papers, \citep{2006A&A...448..623K,  
2006A&A...453..155M}, hereafter Paper I and Paper II, we reported the  
discovery of faint circumstellar envelops (CSE) around Galactic  
classical Cepheids. Interestingly, all the Cepheids we observed  
(\object{$\ell$~Car} in Paper I, \object{$\alpha$~UMi}, and  
\object{$\delta$~Cep} in Paper II) were found to harbor CSE with  
similar characteristics: a CSE 3 to 4 times larger than the star which  
accounts for a few percent of the total flux in the infrared $K$  
band. The presence of CSE was discovered in our attempt to improve our  
knowledge of Cepheids in the context of distance determination via the  
interferometric Baade-Wesselink method (IBWM). Part of the method  
requires the measurement of the angular diameter variation of the star  
during its pulsation. The determination of the angular diameters from  
sparse interferometric measurements is not straightforward because  
optical interferometers gather high angular resolution data only at a  
few baselines at a time, thus good phase and angular resolution  
coverage cannot be achieved in a short time. For Cepheids, the main  
uncertainty in the IBWM was thought to be the center-to-limb darkening  
(CLD), which biases the interferometric angular diameter measurements  
\citep{2004ApJ...603..285M}.  
  
The direct measurement of CLD is possible using an optical  
interferometer, given sufficient angular resolution and  
precision. Among current optical interferometers, CHARA/FLUOR  
\citep{2005ApJ...628..453T, 2006SPIE.6268E..46M} is one of the few  
capable of such a measurement for Cepheids. The only Cepheid  
accessible to CHARA/FLUOR, i.e. large enough in angular diameter, for  
such a measurement is Polaris ($\alpha$~UMi), which we observed and  
found to have a CLD compatible with hydrostatic photospheric  
models, though surrounded by a CSE (Paper II). Polaris, however, is a  
very low amplitude pulsation Cepheid: 0.4\% in diameter, compared to  
15 to 20\% for type I Cepheids \citep{Moskalik2005}, thus the  
agreement is not necessarily expected for large amplitude Cepheids,  
whose photospheres are more out of equilibrium. The direct measurement  
of CLD of a high amplitude Cepheid during its pulsation phase remains  
to be performed.  
  
Hydrodynamic simulations \citep{2003ApJ...589..968M} suggest that the  
CLD variations during the pulsation do not account for more than  
a 0.2\% bias in distance determination in the near infrared using the  
IBWM, where most of the IBWM observing work has been done in recent  
years: the best formal distance determination to date using the IBWM  
is of the order of 1.5\% \citep{2005A&A...438L...9M}.  
  
Whereas the near infrared IBWM seems to be relatively immune to bias
from CLD, the recent discovery of CSEs raises the issue of possible
bias in angular diameter measurements, hence bias in distance
estimations at the 10\% level (Paper II). It is therefore important to
continue the study of CSE around Cepheids. We present here
interferometric observations of the non-pulsating supergiant
\object{$\alpha$~Per} and the low amplitude Cepheid \object{Y~Oph}. We
obtained these results in the near infrared $K$-band, using the Fiber
Linked Unit for Optical Recombination --- FLUOR ---
\citep{2006SPIE.6268E..46M}, installed at Georgia State University's
Center for High Angular Resolution Astronomy (CHARA) Array located on
Mount Wilson, California \citep{2005ApJ...628..453T}.

\section{The low amplitude Cepheid Y~Oph}  

In the General Catalog of Variable Stars \citep{1998GCVS4.C......0K},
Y~Oph is classified in the \texttt{DCEPS} category, i.e. low amplitude
Cepheids with almost symmetrical light curves and with periods less
than 7 days. The GCVS definition adds that \texttt{DCEPS} are first
overtone and/or crossing the instability strip for the first time. A
decrease in photometric variation amplitude over time has been
measured, as well as a period change
\citep{1995AJ....110.1326F}. Using this period change rate,
$7.2\pm1.5\;\mathrm{s} \mathrm{yr}^{-1}$ and the period of
17.1207~days, the star can be identified as crossing the instability
strip for the third time, according to models
\citep{2006PASP..118..410T}.
  
The fact that Y~Oph belongs to the \texttt{DCEPS} category is  
questionable: its period is longer than 7 days, by almost three times,  
though its light curve is quasi-symmetric and with a low amplitude  
compared to other type I Cepheids of similar periods  
\citep{1998MNRAS.296..824V}. Indeed, Y~Oph is almost equally referred  
to in publications as being a fundamental-mode Cepheid or a first  
overtone.  
  
In this context, a direct determination of the linear diameter can  
settle whether Y~Oph belongs to the fundamental mode group or  
not. This is of prime importance: because of its brightness and the  
large amount of observational data available, Y~Oph is often used to  
calibrate the Period-Luminosity (PL) or the Period-Radius (PR)  
relations. The interferometric Baade-Wesselink method offers this  
opportunity to geometrically measure the average linear radius of pulsating  
stars: if Y~Oph is not a fundamental pulsator, its average linear diameter  
should depart from the classical PR relation.

\subsection{Interferometric observations}   
  
The direct detection of angular diameter variations of a pulsating
star has been achieved for many stars now using optical
interferometers \citep{2000Natur.407..485L, 2004A&A...416..941K,
2005A&A...438L...9M}. We showed \citep{2006A&A...453..155M} that for a
given average diameter, one should use a baseline that maximizes the
amplification factor between the variation in angular diameter and
observed squared visibility. This baseline provides an angular
resolution of the order of $B\theta/\lambda \approx 1$, in other words
in the first lobe, just before the first minimum ($B\theta/\lambda
\approx 1.22$ for a uniform disk model), where B is the baseline (in
meters), $\theta$ the angular diameter (in radians) and $\lambda$ the
wavelength of observation (in meters).  According to previous
interferometric measurements \citep{2004A&A...416..941K}, the average
angular diameter of Y~Oph is of the order of 1.45~mas (milli
arcsecond).
    
Ideally, that would mean using a baseline of the order of 300~m, which  
is available at the CHARA Array. Because of a trade we made with other  
observing programs, we used only a 250~m baseline provided by  
telescopes S1 and E2.  
  
The fringes squared visibility is estimated using the integration of  
the fringes power spectrum. A full description of the algorithm can  
be found in \cite{1997A&AS..121..379C} and \cite{2006SPIE.6268E..46M}.  
  
\begin{figure}  
 \plottwo{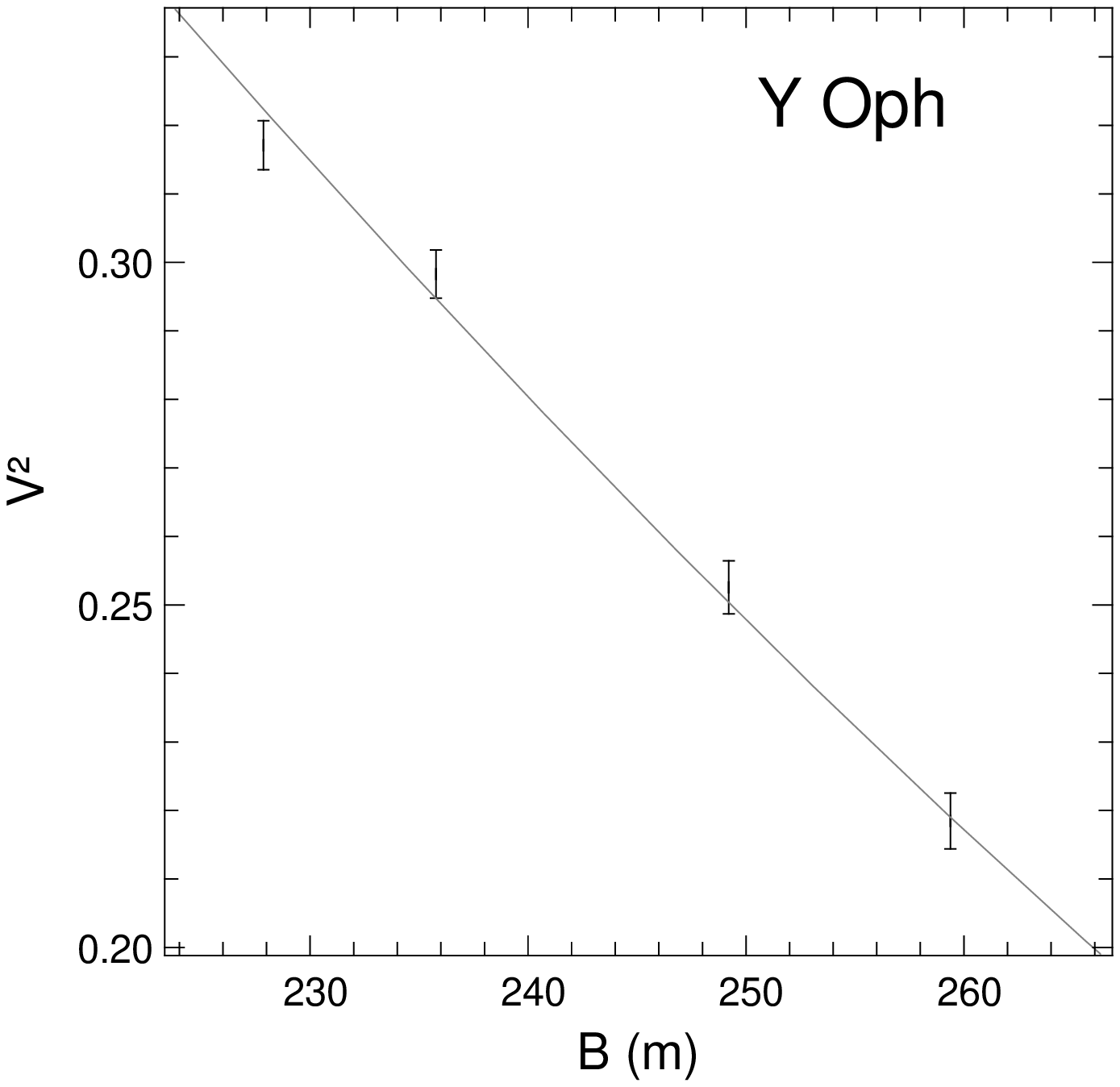}{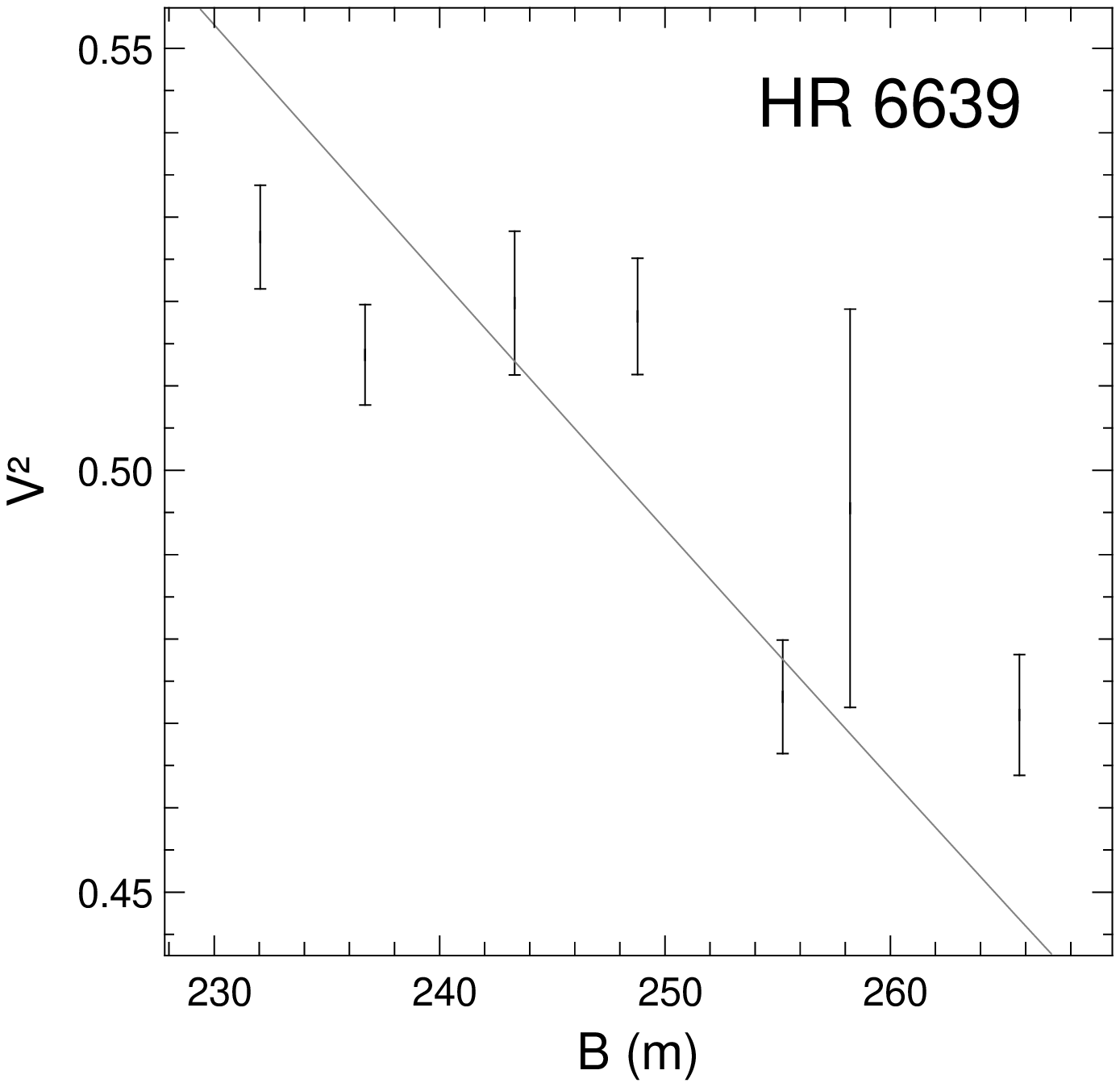}  
 \caption{Evidence that calibrator HR~6639 is probably a binary  
   star. Calibrated squared visibility as a function of baseline for  
   UT 2006/07/10. Left is Y~Oph and right is HR~6639. The  
   interferometric efficiency was supposed constant during the night  
   and established using HR~7809 and $\rho$~Aql with a reduced  
   $\chi^2$ of 1.3. A UD fit, shown as a continuous line, works for  
   Y~Oph ($\chi^2_r=0.6$) whereas it fails for HR~6639  
   ($\chi^2_r=7.2$). This could the sign that HR~6639 has a faint  
   companion: the visibility variation has a 3-5\% amplitude, which  
   corresponds to a magnitude difference of 3 to 4 between the main  
   star and its companion.}  
 \label{y_oph_hr6639}  
\end{figure}

The raw squared visibilities have been calibrated using resolved
calibrator stars, chosen from a specific catalog
\citep{2005A&A...433.1155M} using criteria defined to minimize the
calibration bias and maximize signal to noise. The error introduced by
the uncertainty on each calibrator's estimated angular diameter has
been properly propagated. Among the three main calibrators
(Tab.~\ref{y_oph_calibrators}), one, HR~6639, turned out to be
inconsistent with the others. The raw visibilility of this star was
found to vary too much to be consistent with the expected statistical
dispersion. The quantity to calibrate, the interferometric efficiency
(also called instrument visibility), is very stable for an instrument
using single mode fibers, such as FLUOR. If this quantity is assumed
to be constant over a long period of time, and if observations of a
given simple star are performed several times during this period, one
can check whether or not the variation of the raw visibilities with
respect to the projected baseline is consistent with a uniform disk
model. Doing so, HR~6639 was found inconsistent with the over
stars observed during the same night (Fig~\ref{y_oph_hr6639}). The
unconsistency may be explained by the presence of a faint companion
with a magnitude difference of 3 or 4 with respect to the primary. Two
over calibrators, from another program, were also
used as check stars: HR~7809 and $\rho$~Aql
(Tab.~\ref{y_oph_calibrators}). This latter calibrator is not part of
the catalog by \cite{2005A&A...433.1155M}. Its angular diameter has
been determined using the \cite{2004A&A...426..297K} surface
brightness calibration applied to published photometric data in the
visible and near infrared.
  
For each night we observed Y~Oph, we determined a uniform disk
diameter (Tab.~\ref{Y_Oph_diam}) based on several squared visibility
measurements (Tab.~\ref{Y_Oph_V2}). Each night was assigned a unique
pulsation phase established using the average date of observation and
the \cite{1995AJ....110.1326F} ephemeris, including the measured
period change:
\begin{eqnarray}  
D & = & JD - 2440007.720 \\  
\label{phase}  
E & = & 0.05839D -3.865\times10^{-10}D^2 \\  
P & = & 17.12507 + 3.88\times10^{-6}E  
\end{eqnarray}   
where $E$ is the epoch is the epoch of maximum light (the fractional part is  
the pulsation phase) and $P$ the period at this epoch.  
  
\subsection{Pulsation}  
  
\subsubsection{Radial Velocity integration}  
  
\begin{figure}  
  \plotone{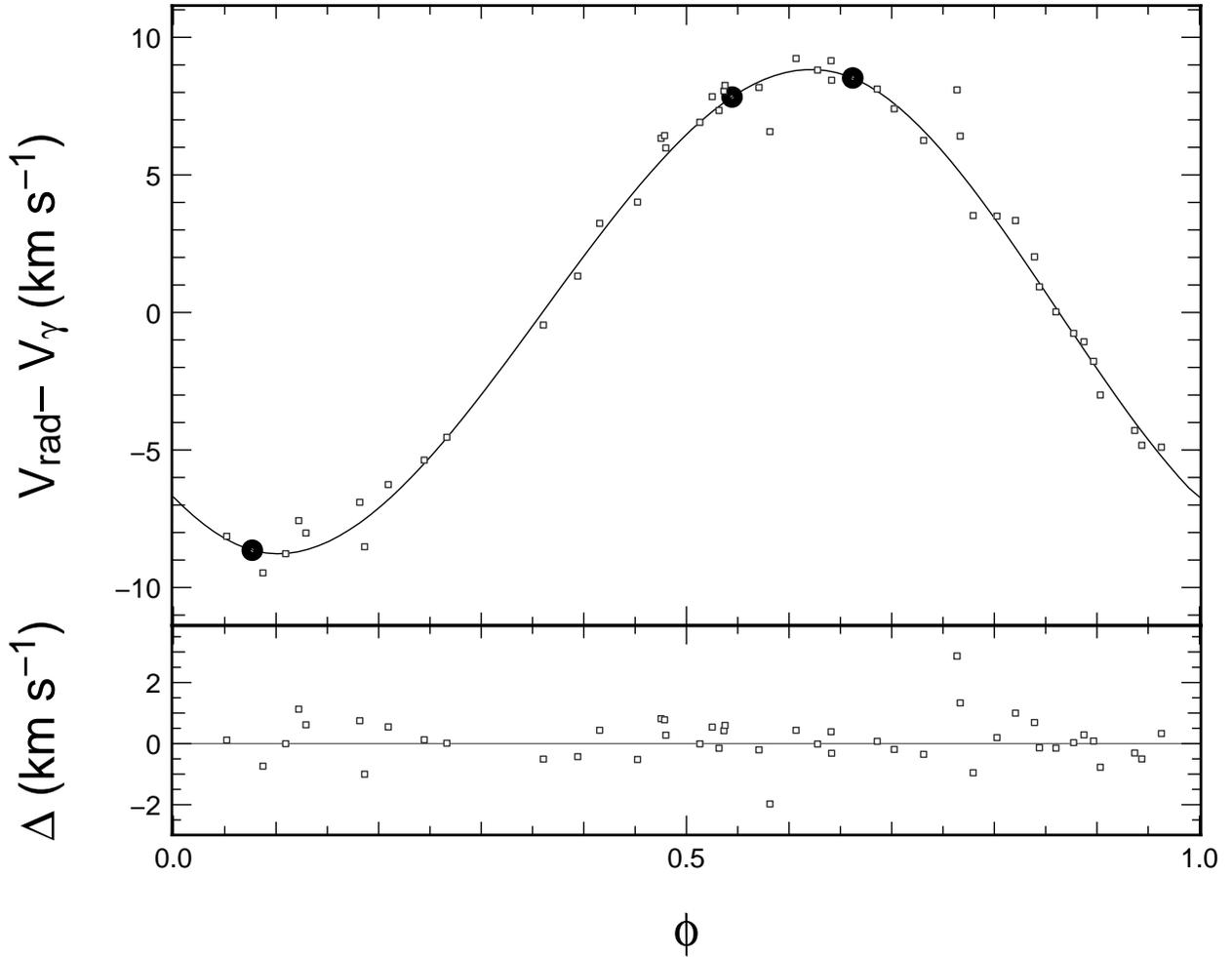}  
  \caption{Y Oph: Radial velocities from \cite{1998AstL...24..815G}.
    The continuous line is the periodic spline function defined by 3
    adjustable floating nodes (large filled circles). The systematic
    velocity, $V_\gamma=-7.9\pm0.1\;\mathrm{km s^{-1}}$, has been
    evaluated using the interpolation function and removed. The lower
    panel displays the residuals of the fit.}
  \label{y_oph_vrad}  
\end{figure}

In order to measure the distance to a pulsating star, the IBWM makes  
use of radial velocities and angular diameters. The latter is the  
integral other time of the former. The radial velocities, which have  
been acquired at irregular intervals during the pulsation phase, must  
be numerically integrated. This process is highly sensitive to noisy  
data and the best way to achieve a robust integration is to  
interpolate the data before integration. For this purpose, we use a  
periodic cubic spline function, defined by floating nodes  
(Fig.~\ref{y_oph_vrad}). The coordinates of these nodes are adjusted  
such that the cubic spline function going through these nodes provides the  
best fit to the data points. The phase positions $\phi_i$ of these  
nodes are forced to be between 0 and 1, but they are replicated every  
$\phi_i+n$, where $n$ is an integer, in order to obtain a periodic  
function of period 1.  
  
Among published Y~Oph radial velocities data, we chose
\cite{1998AstL...24..815G} because of the uniform phase coverage and
the algorithm used to extract radial velocities: the cross-correlation
method. As shown by \cite{2004A&A...428..131N}, the method used can
influence the distance determination \textit{via} the choice of the
so-called projection factor, which we shall introduce in the following
section.  The pulsation phases have been also determined using
Eq.~\ref{phase}.
  
The data presented by \cite{1998AstL...24..815G} were acquired  
between June 1996 and August 1997. As we already mentioned, Y~Oph is  
known for its changing period and photometric amplitude. Based on  
\cite{1995AJ....110.1326F}, the decrease in amplitude observed for the  
photometric B and V bands does not have a measurable counterpart in  
radial velocity. This is why we did not apply any correction in  
amplitude to the radial velocity data in order to take into account  
the ten years between the spectroscopic and interferometric  
measurements.  
  
\subsubsection{Distance determination method}  
  
\begin{figure}  
  \plotone{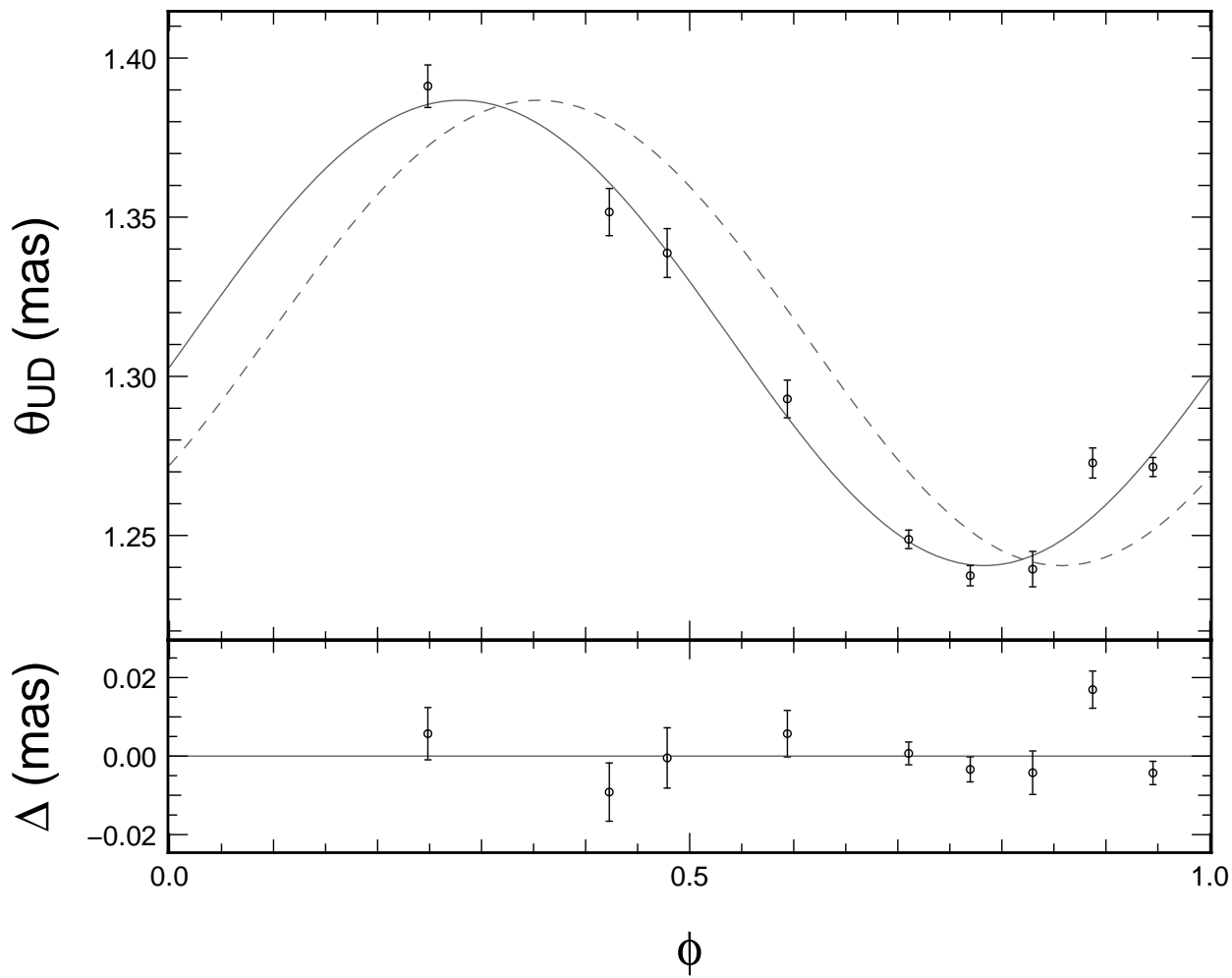}  
  \caption{Y Oph: angular diameter variations. Upper panel:  
    CHARA/FLUOR uniform disk angular diameters as a function of  
    phase. Each data point corresponds to a given night, which  
    contains several individual squared visibility measurements  
    (Tab.~\ref{Y_Oph_diam}). The solid line is the integration of  
    the radial velocity (Fig.~\ref{y_oph_vrad}) with distance, average  
    angular diameter and phase shift adjusted. The dashed line has a  
    phase shift set to zero. The lower panel shows the residuals to  
    the continuous line.}  
  \label{y_oph_ibwm}  
\end{figure}

Once radial velocities $v_\mathrm{rad}$ are interpolated  
(Fig.~\ref{y_oph_vrad}) and integrated, the distance $d$ is determined  
by fitting the radial displacement to the measured angular diameters  
(Fig.~\ref{y_oph_ibwm}):  
			   \begin{equation}  
	   \theta_\mathrm{UD}(T) - \theta_\mathrm{UD}(0) =  
	     -2\frac{kp}{d} \int_0^T v_\mathrm{rad}(t)dt  
			     \label{ibwm}  
			    \end{equation}  
where $\theta_\mathrm{UD}$ is the interferometric uniform disk
diameter, and $k$ is defined as the ratio between $\theta_\mathrm{UD}$
and the true stellar angular diameter. The projection factor, $p$, is
the ratio between the pulsation velocity and the spectroscopically
measured radial velocity. The actual parameters of the fit are the
average angular diameter $\theta_\mathrm{UD}(0)$ and the biased
distance $\frac{d}{kp}$.
  
This formalism assumes that both $k$ and $p$ do not vary during the  
pulsation. There is evidence that this might be true for $k$, based on  
hydrodynamic simulation \citep{2003ApJ...589..968M}, at the $0.2\%$  
level. Observational evidence exists as well: when we measured the  
$p$-factor of $\delta$~Cep \citep{2005A&A...438L...9M} we did not find  
any difference between the shapes of the left and right parts of  
Eq.~\ref{ibwm}, therefore $kp$ is probably constant other a pulsation  
period, at least at the level of precision we have available.  
  
For this work, we will adopt the value for $p$ we determined
observationally for near infrared interferometry/ cross correlation
radial velocity: $p=1.27$. This result has been established for
$\delta$~Cep \citep{2005A&A...438L...9M}. This is also the latest
value computed from hydrodynamical photospheric models
\citep{2004A&A...428..131N}. The IBWM fit yields a biased distance
$d/k=480\pm18\;\mathrm{pc}$ and an average angular uniform disk
diameter $\theta_\mathrm{UD}(0) = 1.314\pm0.005\;\mathrm{mas}$. Note
that we had to allow a phase shift between interferometric and radial
velocity observations: $-0.074\pm0.005$ (Fig.~\ref{y_oph_ibwm}). The
final reduced $\chi^2$ is of the order of 3, mostly due to one data
point ($\phi=0.887$).
  
\subsubsection{Choice of $k$}  
  
Usually, the choice of $k$ is made assuming the star is a
limb-darkened disk. The strength of the CLD is computed using
photospheric models, then a value of $k$ is computed. This approach is
sometimes confusing because, even for a simple limb darkened disk,
there is no unique value of $k$, in the sense that this value varies
with respect to angular resolution. The uniform disk angular size
depends upon which portion of the visibility curve is measured.
However, it is mostly unambiguous in the first lobe of visibility,
i.e. at moderate angular resolution: $B\theta/\lambda \leq 1$.
  
However, as shown in Paper II, the presence of a faint CSE around
Cepheids biases $k$ up to 10\%, particularly when the angular
resolution is moderate and the star is not well resolved ($V^2\sim
0.5$). Under these conditions, the CSE is largely resolved, leading to
a strong bias if the CSE is omitted. On the other hand, at greater
angular resolution ($B\theta/\lambda\sim 1$), the star is fully
resolved ($V^2$ approaches it first null) and the bias from the CSE is
minimized.  In any cases, it is critical to determine whether or not
Y~Oph is surrounded by a CSE if an accurate distance is to be derived.

\subsection{Interferometric evidence of a CSE around Y Oph}

We propose here to compare the uniform disk diameters obtained by
VLTI/VINCI \citep{2004A&A...416..941K} and CHARA/FLUOR (this
work). This makes sense because these two instruments are very
similar.  Both observe in the near infrared $K$-band. Moreover, both
instruments observed Y~Oph in the first lobe of the visibility
profile, though at different baselines.
  
\begin{figure}  
  \plotone{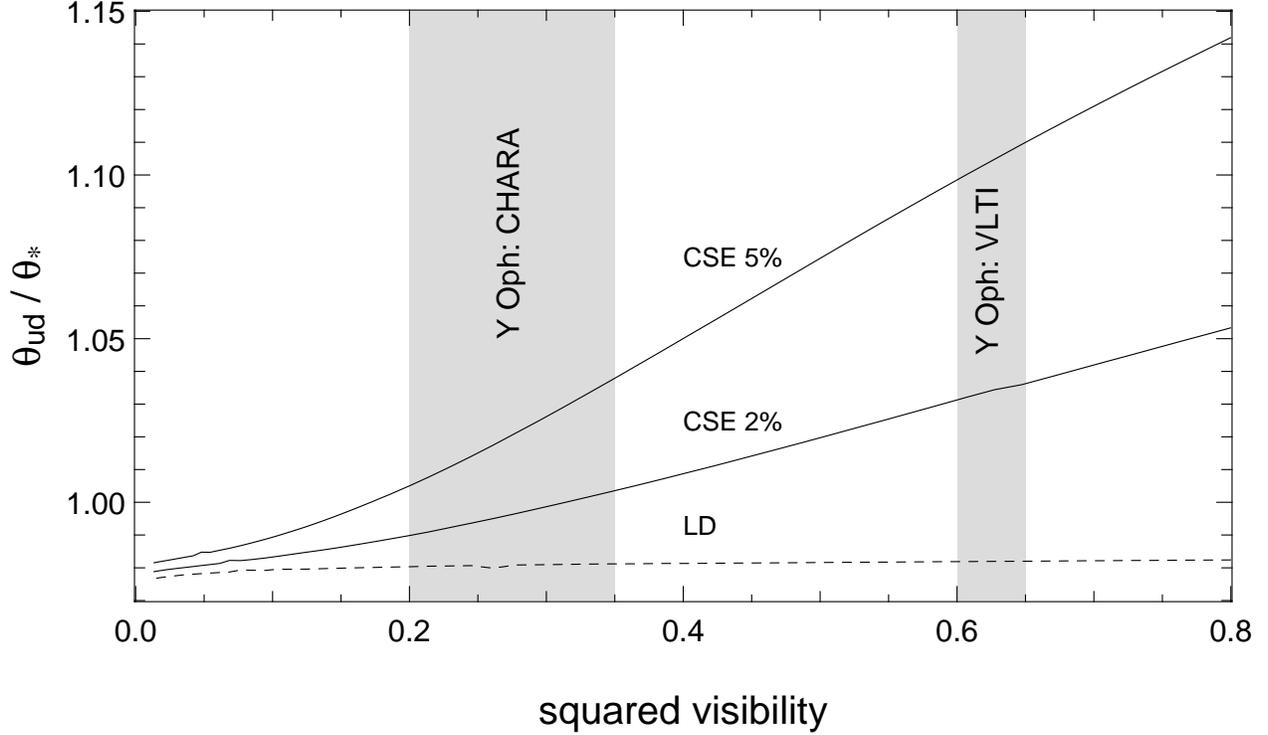}  
  \caption{Y Oph: angular diameter correction factor. We plot here  
    $\theta_\mathrm{UD}/\theta_\star$ for three different Cepheid  
    models as a function of observed squared visibility (first lobe):  
    the dashed line is a simple limb darkened disk with the  
    appropriate CLD strength; the continuous line is a similar LD disk  
    surrounded by a CSE with a 2\% $K$-band flux (short period:  
    Polaris and $\delta$~Cep, see Paper II) and 5\% (long period:  
    $\ell$~Car, see paper I). Note that, in the presence of CSE, the  
    bias is stronger at large visibilities (hence smaller angular  
    resolution). The shaded regions represents near infrared Y~Oph  
    observations: from CHARA (this work) and the VLTI  
    \citep{2004A&A...416..941K}.}  
  \label{y_oph_k}  
\end{figure}

If Y~Oph is truly a uniform disk, or even a limb-darkened disk, the  
two instruments should give similar results. That is because the  
star's first lobe of squared visibility is insensitive to the CLD and  
only dependent the on the size. Conversely, if Y~Oph is  
surrounded by a CSE, we expect a visibility deficit at smaller  
baseline (VLTI/VINCI), hence a larger apparent uniform disk diameter  
(see Fig.~\ref{y_oph_k}). This is because the CSE is fully resolved at  
baselines which barely resolve the star.  
  
\begin{figure}  
  \plotone{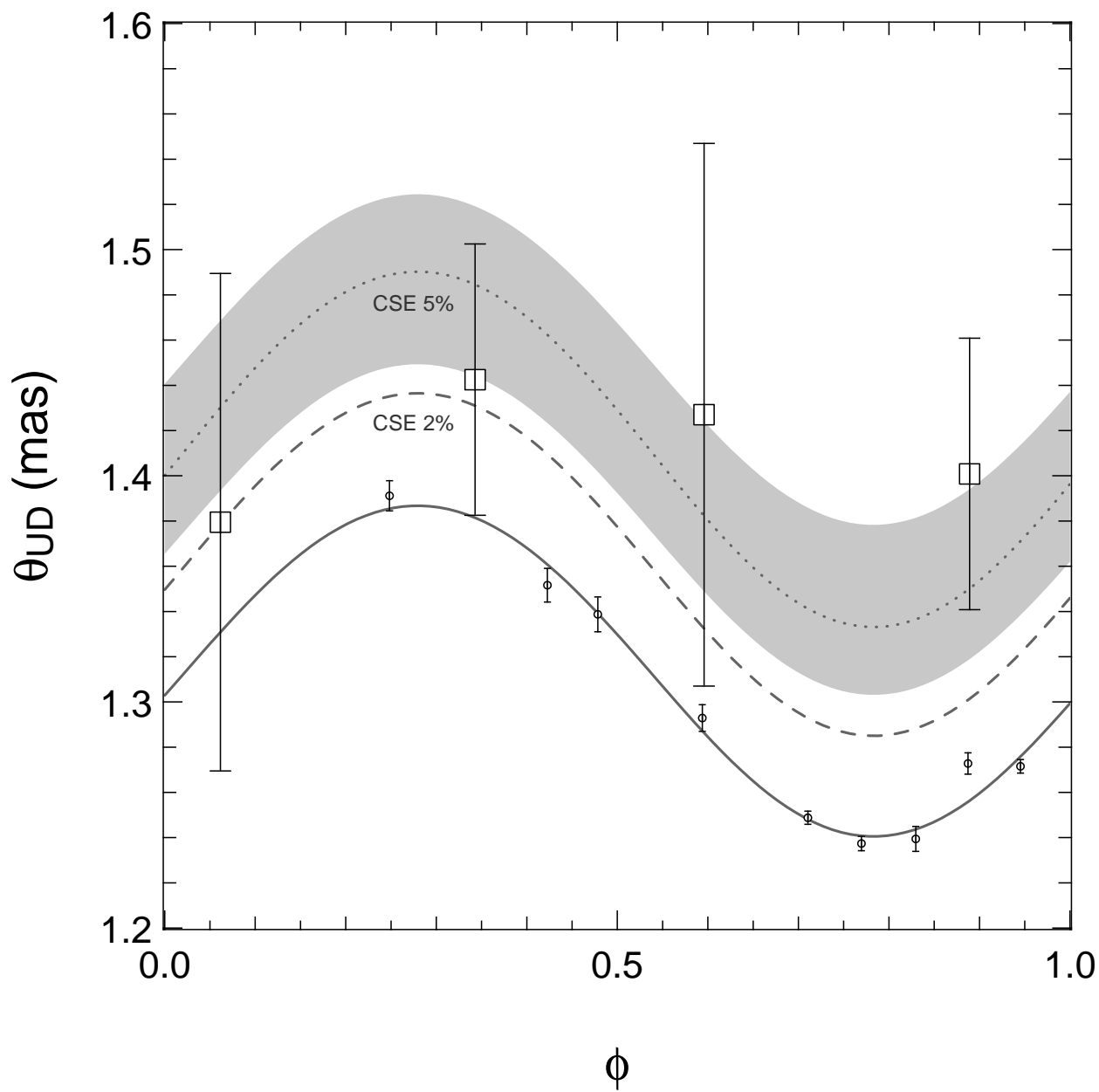}  
  \caption{Y Oph: Comparison between CHARA/FLUOR and VLTI/VINCI
    observations. Uniform disk angular diameter as a function of
    phase. Small data points (with small error bars) and lower
    continuous line: CHARA/FLUOR observations and distance fit. Large
    open squares: VLTI/VINCI observations. The distance fit and it
    uncertainty are reprensenteb by the shaded band. Dashed and dotted
    lines: VLTI/VINCI expected biased observations based on
    CHARA/FLUOR and our CSE model with a flux ratio of 2\% (dashed)
    and 5\% (dotted).}
  \label{y_oph_cse}  
\end{figure}

This is indeed the case, as seen on Fig.~\ref{y_oph_cse}: VLTI/VINCI
UD diameters are larger than CHARA/FLUOR's. Even if the angular
resolution of VLTI/VINCI is smaller than for CHARA/FLUOR, leading to
less precise angular diameter estimations, the disagreement is still
statistically significant, of the order of 3 sigmas. Using the CSE
model we can predict the correct differential UD size between the two
instruments, consistant with the presence of a CSE around Y~Oph. The
amount of discrepancy can be used to estimate the flux ratio between
the CSE and the star. In the case of Y~Oph, we find that the CSE
amounts for $5\pm2\%$ of the stellar flux. Note that for this
comparison, we recomputed the phase of VLTI/VINCI data using
\cite{1995AJ....110.1326F} ephemeris presented in Eq.~\ref{phase}.
  
In Fig.~\ref{y_oph_k}, we plot $k$ as a function of the observed  
squared visibility for different models: hydrostatic CLD, 2\% CSE and  
5\% CSE ($K$-Band flux ratio). For the hydrostatic model we have  
$k=0.983$. For the 5\% CSE models, for CHARA/FLUOR ($0.20<V^2<0.35$),  
$\theta_\mathrm{UD}/\theta_\star = 1.023$. This is the value we shall  
adopt. If we ignore the presence of the CSE, the bias introduced is  
$1.023/0.983\approx1.04$, or 4\%.  
  
  
\subsection{Unbiased distance and linear radius}  
  
This presence of a $5\%$ CSE leads to an unbiased distance of  
$d=491\pm18\;\mathrm{pc}$, which corresponds to a 3.5\% uncertainty on  
the distance. This is to be compared with the bias we corrected for if  
one omits the CSE, of the order of 4\%. Ignoring the CSE leads to a  
distance of $d=472\pm18\;\mathrm{pc}$  
  
We note that $k$ biases only the distance, so one can form  
the following quantity: $[\theta_\mathrm{UD}(0)]\times[d/k]$, which is  
the product of the two adjusted parameters in the fit, both  
biased. This quantity is by definition the linear diameter of the  
star, and does not depend on the factor $k$, even if it is still  
biased by the choice of $p$. If $\theta$ is in mas and $d$ in parsecs,  
then the average linear radius in solar radii is: $R=0.1075\theta d$. In the  
case of Y~Oph, this leads to $R=67.8\pm2.5\;R_\odot$.

\subsection{Conclusion}  
  

Y Oph appears larger (over 2 sigma) in the infrared K-band at a
140 m baseline comapred to a 230 m baseline. Using a model of a star
surrounded by a CSE we developed based on observations of other
Cepheids, this disagreement is explained both qualitatively and
quantitatively by a CSE accounting of 5\% of the stellar flux in the
near infrared $K$-Band. This model allows us to unbias the distance
estimate: $d=491\pm18\;\mathrm{pc}$. The linear radius estimate is not
biased by the presence of CSE and we found $R=67.8\pm2.5\;R_\odot$.
  
Our distance is not consistent with the estimation using the
Barnes-Evans method: \cite{2005ApJ...631..572B} found $d=590\pm42$~pc
(Bayesian) and $d=573\pm8$~pc (least squares). For this work, they
used the same set of radial velocities we used. Our estimate is even
more inconsistent with the other available interferometric estimate,
by \cite{2004A&A...416..941K}: $d=690\pm50\;\mathrm{pc}$. This later
result has been established using an hybrid version of the BW method:
a value of the linear radius is estimated using the Period-Radius
relation calibrated using classical Cepheids, not measured from the
data. This assumption is questionable, as we noted before, since Y~Oph
is a low amplitude Cepheid. \cite{2004A&A...416..941K} deduced
$R=100\pm8\;R_\odot$ from PR relation, whereas we measured
$R=67.8\pm2.5\;R_\odot$. Because Y~Oph's measured linear radius is not
consistent with the PR relation for classical, fundamental mode
Cepheids, it is probably safe to exclude it from further
calibrations. Interestingly, \cite{2005ApJ...631..572B}
observationally determined Y~Oph's linear radius to be also slightly
larger (2.5 sigma) than what we find: $R=92.1\pm6.6\; R_\odot$
(Bayesian) and $R=89.5\pm1.2\;R_\odot$ (least squares). They use a
surface brightness technique using a visible-near infrared color, such
as V-K. This method is biased if the reddening is not well known. If
the star is reddened, V magnitudes are increased more than $K$-band
magnitudes. This leads to an underestimated surface brightness,
because the star appears redder, thus cooler than it is. The total
brightness (estimated from V) is also underestimated. These two
underestimates have opposing effects on the estimated angular
diameter: an underestimated surface brigthness leads to an
overestimated angular diameter, whereas an underestimated luminosity
leads to an underestimated angular diameter. In the case of a
reddening law, the two effects combine and give rise to a larger
angular diameter: the surface brightness effect wins over the total
luminosity. Based on their angular diameter,
$\theta\approx1.45\;\mathrm{mas}$, it appears that
\cite{2005ApJ...631..572B} overestimated Y~Oph's angular size.
Among Cepheids brighter than $m_V=6.0$, Y~Oph has the
strongest B-V color excess, $E(B-V)=0.645$ \citep{1995IBVS.4148....1F}
and one of the highest fraction of polarized light,
$p=1.34\pm0.12\;\%$ \citep{2000AJ....119..923H}. Indeed, Y~Oph is
within the galactic plane: this means that it has a probably a large
extinction due to the interstellar medium.

\section{The non-pulsating yellow super giant $\alpha$ Per}  
  
$\alpha$~Per (HR~1017, HD~20902, F5Ib) is among the most luminous
non-pulsating stars inside the Cepheids' instability strip.  The
Doppler velocity variability has been found to be very weak, of the
order of 100~m/s in amplitude \citep{1998ApJ...494..342B}. This
amplitude is ten times less than what is observed for the very low
amplitude Cepheid Polaris \citep{2000AJ....120..979H}.
  
$\alpha$~Per's apparent angular size, approximately 3 mas
\citep{2001AJ....122.2707N}, makes it a perfect candidate for direct
center-to-limb darkening detection with CHARA/FLUOR. Following the
approach we used for Polaris (Paper II), we observed $\alpha$~Per
using three different baselines, including one sampling the second
lobe, in order to measure its CLD strength, but also in order to be
able to assess the possible presence of a CSE around this star.

\subsection{Interferometric observations}  
  
If the star is purely a CLD disk, then only two baselines are required  
to measure the angular diameter and the CLD strength. Observations  
must take place in the first lobe of the squared visibility profile,  
in order to set the star's angular diameter $\theta$. The first lobe  
is defined by $B\theta_\mathrm{UD}/\lambda < 1.22$.  Additional  
observations should taken in the second lobe, in particular near the  
local maximum ($B\theta/\lambda \sim 3/2$), because the strength of  
the CLD is directly linked to the height of the second lobe. To  
address the presence of a CSE, observations should be made at a small  
baseline. Because the CSEs that were found around Cepheids are roughly 3  
times larger that the star itself (Paper I and II), we chose a small  
baseline where $B\theta/\lambda \sim 1/3$.  As demonstrated by our  
Polaris measurements, the presence of CSE is expected to weaken the  
second lobe of visibility curve, mimicking stronger CLD.  
  
\begin{figure}  
  \plotone{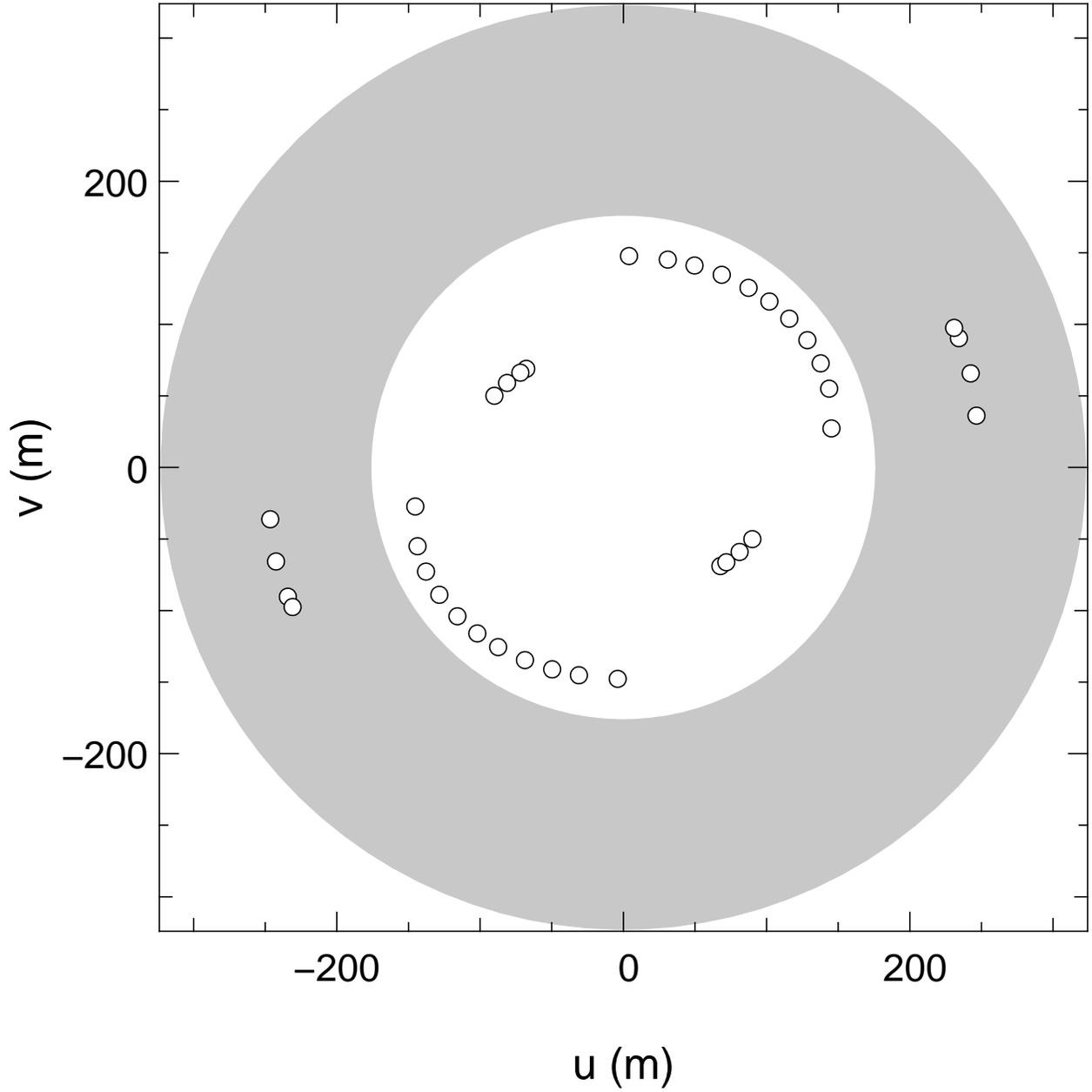}  
    \caption{$\alpha$~Per. Projected baselines, in meters.  North is up   
      East is right. The shaded area corresponds to the squared  
      visibility's second lobe. The baselines are W1-W2, E2-W2 and E2-W1,   
      from the shortest to the longest.}  
    \label{mirfak_uv}  
\end{figure}

FLUOR operates in the near infrared $K$-band, with a mean wavelength  
of $\lambda_0\approx2.13\;\mu\mathrm{m}$. This sets the small,  
first-lobe and second-lobe baselines at approximatively 50, 150 and  
220 meters, which are well-matched to CHARA baselines W1-W2  
($\sim100$~m), E2-W2 ($\sim150$~m) and E2-W1 ($\sim250$~m). See  
Fig.~\ref{mirfak_uv} for a graphical representation of the baseline  
coverage.  
  
The data reduction was performed using the same pipeline as for Y~Oph.  
Squared visibilities were calibrated using a similar strategy we  
adopted for Y~Oph. We used two sets of calibrators: one for the  
shorter baselines, W1-W2 and E2-W2, and one for the longest baseline,  
E2-W1 (Tab.~\ref{mirfak_calibrators}).  
    
\subsection{Simple Model}  
  
\subsubsection{Limb darkened disk}  
  
To probe the shape of the measured visibility profile, we first used  
an analytical model which includes the stellar diameter and a CLD  
law. Because of its versatility, we adopt here a power law  
\citep{1997A&A...327..199H}: $I(\mu)/I(0) = \mu^\alpha$, with $\mu =  
\cos(\theta)$, where $\theta$ is the angle between the stellar surface  
and the line of sight. The uniform disk diameter case corresponds to  
$\alpha=0.0$, whereas an increasing value of $\alpha$ corresponds to a  
stronger CLD.  
  
All visibility models for this work have been computed taking into
account instrumental bandwidth smearing
\citep{2006ApJ...645..664A}. From this two-parameter fit, we deduce a
stellar diameter of $\theta_\alpha = 3.130\pm0.007$ mas and a power
law coefficient $\alpha = 0.119\pm0.016$. The fit yields a reduced
$\chi^2 = 1.0$. There is a strong correlation between the diameter and
CLD coefficent: 0.9. This is a well known effect, that a change in CLD
induces a change in apparent diameter.
  
\begin{figure}  
  \plotone{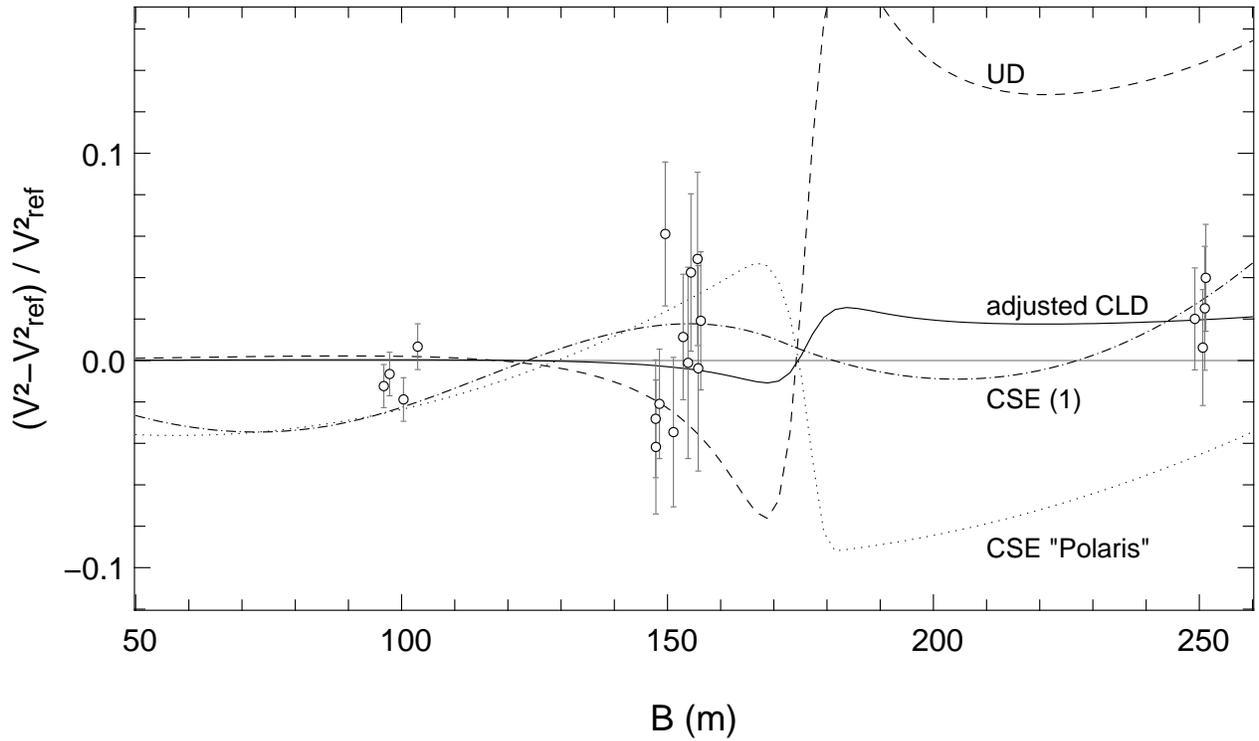}  
  \caption{$\alpha$~Per squared visibility models: UD, adjusted CLD,  
    PHOENIX and 2 different CSE models are plotted here as the  
    residuals to the PHOENIX CLD with respect to baseline. The common  
    point at $B\approx175$~m is the first minimum of the visibility  
    function. The top of the second lobe is reached for  
    $B\approx230$~m. See Tab.~\ref{mirfak_models} for parameters and  
    reduced $\chi^2$ of each model.}  
  \label{mirfak_cse}  
\end{figure}

This fit is entirely satisfactory, as the reduced $\chi^2$ is of the  
order unity and the residuals of the fit do not show any trend  
(Fig.~\ref{mirfak_cse}). We note that this is not the case for Polaris  
(Paper II). Polaris, with very similar u-v coverage to $\alpha$~Per,  
does not follow a simple LD disk model because it is surrounded by a  
faint CSE.  
  
\subsubsection{Hydrostatic models}  
  
We computed a small grid of models which consists of six  
one-dimensional, spherical, hydrostatic models using the PHOENIX  
general-purpose stellar and planetary atmosphere code version 13.11;  
for a description see \cite{1999ApJ...525..871H} and  
\cite{1999JCAM_109_41H}.  The range of effective temperatures and  
surface gravities is based on a summary of $\alpha$~Per's parameters  
by \cite{1996AJ....111.2099E}:    
\begin{itemize}  
\item effective temperatures, $T_{\rm eff}$ = 6150~K, 6270~K, 6390~K;  
\item $\log(g) = 1.4, 1.7$;  
\item radius, $R =3.92\times 10^{12}$ cm;  
\item depth-independent micro-turbulence, $\xi$ = 4.0 km s$^{-1}$;  
\item mixing-length to pressure scale ratio, 1.5;  
\item solar elemental abundance LTE atomic and molecular line  
  blanketing, typically $10^6$ atomic lines and $3\times 10^{5}$  
  molecular lines dynamically selected at run time;  
\item non-LTE line blanketing of \ion{H}{1} (30 levels, 435  
  bound-bound transitions), \ion{He}{1} (19 levels, 171 bound-bound  
  transitions), and \ion{He}{2} (10 levels, 45 bound-bound transitions);  
\item boundary conditions: outer pressure, $10^{-4}$ dynes cm$^{-2}$,  
  extinction optical depth at 1.2~$\mu \mathrm{m}$: outer $10^{-10}$,  
  inner $10^{2}$.  
\end{itemize}  
For this grid of models the atmospheric structure is computed at 50  
radial shells (depths) and the radiative transfer is computed along 50  
rays tangent to these shells and 14 additional rays which impact the  
innermost shell, the so-called core-intersecting rays. The  
intersection of the rays with the outermost radial shell describes a  
center-to-limb intensity profile with 64 angular points.  
    
Power law fits to the hydrostatic model visibilities range from  
$\alpha=0.132$ to $\alpha=0.137$ (Tab.~\ref{phoenix_models}), which  
correspond to 0.8 to 1.1 sigma above the value we measured. Our  
measured CLD is below that predicted by the the models, or in other  
words the predicted darkening is slightly overestimated.  
  
\subsection{Possible presence of CSE}  
  
Firstly, we employ the CSE model used for Cepheids (Paper I, Paper  
II). This model is purely geometrical and it consists of a limb darkened  
disk surrounded by a faint CSE whose two-dimensional geometric  
projection (on the sky) is modeled by a ring. The ring parameters  
consist of the angular diameter, width and flux ratio with respect to  
the star ($F_s/F_\star$). We adopt the same geometry found for  
Cepheids: a ring with a diameter equal to 2.6 times the stellar  
diameter (Paper II) with no dependency with respect to the width  
(fixed to a small fraction of the CSE angular diameter, say 1/5).  
  
This model restriction is justified because testing the agreement  
between a generic CSE model and interferometric data requires a  
complete angular resolution coverage, from very short to very large  
baselines. Though our $\alpha$~Per data set is diverse regarding the  
baseline coverage, we lack a very short baseline ($B\sim50$\,m) which  
was not available at that time at the CHARA Array.   
  
The simplest fit consists in adjusting the geometrical scale: the  
angular size ratio between the CSE and the star is fixed. This yields  
a reduced $\chi^2$ of 3, compared to 1.1 for a hydrostatic LD and 1.0  
for an adjusted CLD law (Tab.~\ref{mirfak_models}).  
  
We can force the data to fit a CSE model by relaxing the CLD of the
star or the CSE flux ratio. A fit of the size of the star and the
brightness of the CSE leads to a reduced $\chi^2$ of 1.4 and results
in a very small flux ratio between the CSE and the star of
$0.0006\pm0.0026$, which provides an upper limit for the CSE flux of
0.26\% (compared to 1.5\% for Polaris and $\delta$~Cep for example,
and 5\% for Y~Oph). On the other hand, forcing the flux ratio to match
that for Cepheids with CSEs and leaving the CLD free leads to a
reduced $\chi^2$ of 1.4 but also to a highly unrealistic
$\alpha=0.066\pm0.004$ (Tab.~\ref{mirfak_models} and
Fig.~\ref{mirfak_cse}).
  
This leads to the conclusion that the presence of a CSE around  
$\alpha$~Per similar to the measured Cepheid CSE is highly  
improbable. As shown above, the measured CLD is slightly weaker (1  
sigma) than one predicted by model atmospheres. A compatible CSE model  
exists if the CLD is actually even weaker but in unrealistic proportions.  
  
\subsection{Conclusion}  
  
The observed $\alpha$~Per visibilities are not compatible with the  
presence of a CSE similar to that detected around Cepheids. The data  
are well explained by an adjusted center-to-limb darkening. The  
strength of this CLD is compatible with the hydrostatic model within one  
sigma.  
    
\section{Discussion}  
  
Using the interferometric Baade-Wesselink method, we determined the
distance to the low amplitude Cepheid Y~Oph to be
$d=491\pm18$~pc. This distance has been unbiased from the presence of
the circumstellar envelope. This bias was found to be of the order of
4\% for our particular angular resolution and the amount of CSE we
measured. Y~Oph's average linear diameter has also been estimated to
be $R=67.8\pm2.5\;R_\odot$. This latter quantity in intrinsically
unbiased by the center-to-limb darkening or the presence of a
circumstellar envelope. This value is found to be substantially lower,
almost 4 sigmas, than the Period-Radius relation prediction:
$R=100\pm8\;R_\odot$. Among other peculiarities, we found a very large
phase shift between radial velocity measurements and interferometric
measurements: $\Delta\phi=-0.074\pm0.005$, which corresponds to more
that one day.  For these reasons, we think Y~Oph should be excluded
from future calibrations of popular relations for classical Cepheids,
such as Period-Radius, Period-Luminosity relations.

So far, the four Cepheids we looked at have a CSE: $\ell$~Car (Paper I),  
Polaris, $\delta$~Cep (Paper II) and Y~Oph (this work). On the other  
hand, we have presented here similar observations of $\alpha$~Per, a  
non-pulsating supergiant inside the instability strip, which provides  
no evidence for a circumstellar envelope. This non detection reinforces  
the confidence we have in our CSE detection technique and draws more  
attention to a pulsation driven mass loss mechanism.  
  
Interestingly, there seems to be a correlation between the period and
the CSE flux ratio in the $K$-Band: short-period Cepheids seem to have
a fainter CSE compared to longer-period Cepheids
(Tab.~\ref{cepheids_CSE}).  Cepheids with long periods have higher
masses and larger radii, thus, if we assume that the CSE $K$-Band
brightness is an indicator of the mass loss rate, this would mean that
heavier stars experience higher mass loss rates. This is of prime
importance in the context of calibrating relations between Cepheids'
fundamental parameters with respect to their pulsation periods. If
CSEs have some effects on the observational estimation of these
fundamental parameters (luminosity, mass, radius, etc.), a correlation
between the period and the relative flux of the CSE can lead to a
biased calibration.

\begin{acknowledgments}  
  Research conducted at the CHARA Array is supported at Georgia State
  University by the offices of the Dean of the College of Arts and
  Sciences and the Vice President for Research. Additional support for
  this work has been provided by the National Science Foundation through
  grants AST 03-07562 and 06-06958. We also wish to acknowledge the
  support received from the W.M. Keck Foundation. This research has made
  use of SIMBAD database and the VizieR catalogue access tool, operated
  at CDS, Strasbourg, France.
\end{acknowledgments}

  
\clearpage


\clearpage  
  
\begin{figure}  
  \plotone{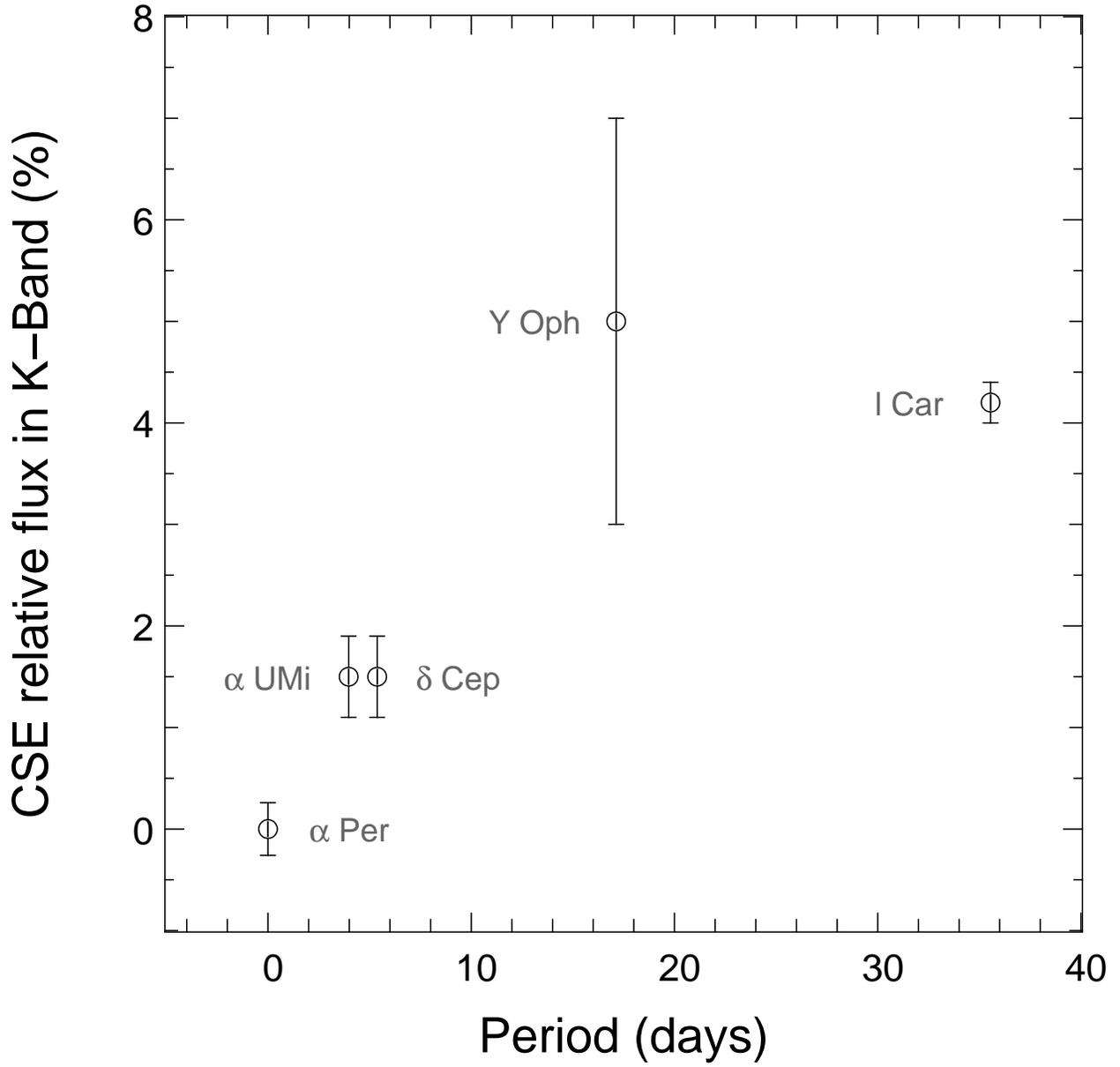}  
  \caption{measured relative $K$-band CSE fluxes (in percent) around
    Cepheids, as a function of the pulsation period (in days). The
    non-pulsating yellow supergiant $\alpha$~Per is plotted with P=0.}
  \label{cepheids_cse_P}  
\end{figure}  
  
\clearpage  
 
\begin{deluxetable}{lcccc}  
  \tablecaption{Y~Oph calibrators.\label{y_oph_calibrators}} 
  \tablehead{ \colhead{Star} &  \colhead{S.T.} & 
    \colhead{UD Diam.} & \colhead{Notes} \\ & & (mas) }  
  \startdata  
  \object{HD 153033} & K5III & $1.100\pm0.015$ & \\  
  \object{HD 175583} & K2III & $1.021\pm0.014$ & \\  
  \object{HR 6639}   & K0III & $0.904\pm0.012$ & rejected\\  
  \hline  
  \object{HR 7809}   & K1III & $1.055\pm0.015$ & \\  
  \object{$\rho$~Aql} & A2V  & $0.370\pm0.005$ & not in M05
  \enddata  
  \tablecomments{``S.T.''  
    stands for spectral type. Uniform Disk diameters, given in mas,  
    are only intended for computing the expected squared visibility in  
    the $K$-band. All stars but $\rho$ Aql are from M05 catalog  
    \citep{2005A&A...433.1155M}. Refer to text for an explanation why  
    HR~6639 has been rejected.}  
\end{deluxetable}  
  
\clearpage  
  
\begin{deluxetable}{lrrc}  
    \tablecaption{Journal of observations: Y~Oph. \label{Y_Oph_V2}}  
  \tablehead{  
    \colhead{MJD-53900} & \colhead{B} & \colhead{P.A.} &  
    \colhead{$V^2$} \\ & (m) & (deg)}  
    \startdata  
    18.260 & 229.808 &  63.540 & $ 0.2348\pm 0.0061$ \\ 
    18.296 & 215.049 &  71.242 & $ 0.2823\pm 0.0065$ \\ 
    18.321 & 207.247 &  77.913 & $ 0.3030\pm 0.0075$ \\ 
    21.245 & 232.792 &  62.353 & $ 0.2511\pm 0.0065$ \\ 
    21.283 & 216.576 &  70.243 & $ 0.2954\pm 0.0085$ \\ 
    21.309 & 208.313 &  76.765 & $ 0.3194\pm 0.0085$ \\ 
    22.212 & 245.894 &  58.023 & $ 0.2092\pm 0.0071$ \\ 
    22.234 & 236.338 &  61.050 & $ 0.2352\pm 0.0078$ \\ 
    22.251 & 228.941 &  63.902 & $ 0.2615\pm 0.0085$ \\ 
    24.188 & 253.988 &  55.914 & $ 0.2108\pm 0.0062$ \\ 
    24.212 & 243.641 &  58.680 & $ 0.2427\pm 0.0060$ \\ 
    24.231 & 235.388 &  61.389 & $ 0.2634\pm 0.0065$ \\ 
    26.168 & 259.377 &  54.714 & $ 0.2185\pm 0.0041$ \\ 
    26.194 & 249.210 &  57.113 & $ 0.2526\pm 0.0039$ \\ 
    26.224 & 235.772 &  61.251 & $ 0.2983\pm 0.0034$ \\ 
    26.242 & 227.859 &  64.367 & $ 0.3171\pm 0.0035$ \\ 
    27.194 & 247.788 &  57.495 & $ 0.2650\pm 0.0040$ \\ 
    27.211 & 240.340 &  59.704 & $ 0.2854\pm 0.0038$ \\ 
    27.228 & 232.806 &  62.348 & $ 0.3177\pm 0.0045$ \\ 
    27.250 & 223.415 &  66.429 & $ 0.3385\pm 0.0045$ \\ 
    28.241 & 225.978 &  65.207 & $ 0.3332\pm 0.0045$ \\ 
    28.257 & 219.413 &  68.547 & $ 0.3629\pm 0.0061$ \\ 
    29.219 & 234.329 &  61.775 & $ 0.2857\pm 0.0049$ \\ 
    29.237 & 226.405 &  65.012 & $ 0.3148\pm 0.0053$ \\ 
    29.253 & 219.709 &  68.380 & $ 0.3320\pm 0.0054$ \\ 
    30.192 & 245.267 &  58.203 & $ 0.2510\pm 0.0035$ \\ 
    30.216 & 234.738 &  61.624 & $ 0.2893\pm 0.0041$ \\ 
    30.235 & 226.418 &  65.007 & $ 0.3112\pm 0.0044$ \\ 
    30.244 & 222.544 &  66.866 & $ 0.3226\pm 0.0046$ \\ 
    30.261 & 215.937 &  70.653 & $ 0.3462\pm 0.0048$ \\ 
    \enddata  
    \tablecomments{Date of observation (modified julian day), telescope  
      projected separation (m), baseline projection angle (degrees)  
      and squared visibility}  
\end{deluxetable}

\clearpage  

\begin{deluxetable}{cccccc}  
 \tablecaption{Y~Oph angular diameters.\label{Y_Oph_diam}}  
 \tablehead{  
   \colhead{MJD-53900} & \colhead{$\phi$} &  
   \colhead{$N_\mathrm{obs.}$} & \colhead{B}  &  
   \colhead{$\theta_\mathrm{UD}$} & \colhead{$\chi^2$} \\
   & & (m) & (mas) }  
 \startdata  
 18.292 & 0.248 & 3 & 207-230 & $1.3912\pm0.0067$ & 0.55\\   
 21.279 & 0.423 & 3 & 208-233 & $1.3516\pm0.0074$ & 1.06\\   
 22.232 & 0.478 & 3 & 229-246 & $1.3387\pm0.0077$ & 0.05\\   
 24.210 & 0.594 & 3 & 235-254 & $1.2929\pm0.0059$ & 0.19\\   
 26.207 & 0.710 & 4 & 228-259 & $1.2488\pm0.0029$ & 0.61\\   
 27.221 & 0.770 & 4 & 223-248 & $1.2374\pm0.0032$ & 0.76\\   
 28.249 & 0.830 & 2 & 219-226 & $1.2394\pm0.0055$ & 0.30\\   
 29.237 & 0.887 & 3 & 220-234 & $1.2728\pm0.0047$ & 0.26\\   
 30.229 & 0.945 & 5 & 216-245 & $1.2715\pm0.0030$ & 0.56
 \enddata  
 \tablecomments{Data points have been  
   reduced to one uniform disk diameter per night. Avgerage date of  
   observation (modified julian day), pulsation phase, number of  
   calibrated $V^2$, projected baseline range, uniform disk  
   angular diameter and reduced $\chi^2$.}  
\end{deluxetable}

\clearpage  

\begin{deluxetable}{lllc}  
  \tablecaption{$\alpha$~Per calibrators.\label{mirfak_calibrators}}  
   \tablehead{  
    \colhead{Name} & \colhead{S.T.} & \colhead{UD Diam.} &   
    \colhead{Baselines}\\
    & & (mas)    }  
   \startdata  
   \object{HD 18970}  & G9.5III  & $1.551\pm0.021$ & W1-W2, E2-W2\\  
   \object{HD 20762}  & K0II-III & $0.881\pm0.012$ & E2-W1\\  
   \object{HD 22427}  & K2III-IV & $0.913\pm0.013$ & E2-W1\\  
   \object{HD 31579}  & K3III    & $1.517\pm0.021$ & W1-W2, E2-W2\\  
   \object{HD 214995} & K0III    & $0.947\pm0.013$ & W1-W2 
   \enddata  
   \tablecomments{``S.T.'' stands for spectral  
     type. Uniform Disk diameters, given in mas, are only intended for  
     computing the expected squared visibility in the $K$-band. These  
     stars come from our catalog of interferometric calibrators  
     \citep{2005A&A...433.1155M}}  
\end{deluxetable}

\clearpage  
  
\begin{deluxetable}{lrrc}  
    \tablecaption{Journal of observations: $\alpha$~Per. \label{mirfak_V2}}  
    \tablehead{  
      \colhead{MJD-54000} & \colhead{B} & \colhead{P.A.} &  
      \colhead{$V^2$} \\ & (m) & (deg)}  
    \startdata  
    46.281 & 147.82  &  79.37 & $ 0.02350\pm 0.00075$ \\   
    46.321 & 153.87  &  69.04 & $ 0.01368\pm 0.00067$ \\   
    46.347 & 155.79  &  62.17 & $ 0.01093\pm 0.00060$ \\   
    46.372 & 156.27  &  55.30 & $ 0.01134\pm 0.00039$ \\   
    46.398 & 155.65  &  48.10 & $ 0.01197\pm 0.00051$ \\   
    46.421 & 154.41  &  41.32 & $ 0.01367\pm 0.00052$ \\   
    46.442 & 152.94  &  34.84 & $ 0.01581\pm 0.00045$ \\   
    46.466 & 151.13  &  27.03 & $ 0.01824\pm 0.00064$ \\   
    46.488 & 149.58  &  19.40 & $ 0.02167\pm 0.00071$ \\   
    46.510 & 148.49  &  12.06 & $ 0.02250\pm 0.00058$ \\   
    46.539 & 147.77  &   1.53 & $ 0.02370\pm 0.00070$ \\   
    47.225 &  96.66  & -44.51 & $ 0.27974\pm 0.00282$ \\   
    47.233 &  97.75  & -47.34 & $ 0.27321\pm 0.00274$ \\   
    47.252 & 100.37  & -53.97 & $ 0.24858\pm 0.00258$ \\   
    47.274 & 103.03  & -60.91 & $ 0.23529\pm 0.00242$ \\   
    47.327 & 249.17  &  81.65 & $ 0.01363\pm 0.00032$ \\   
    47.352 & 251.22  &  74.84 & $ 0.01340\pm 0.00032$ \\   
    47.374 & 251.04  &  68.91 & $ 0.01330\pm 0.00035$ \\   
    47.380 & 250.67  &  67.10 & $ 0.01316\pm 0.00035$    
    \enddata  
    \tablecomments{Date of observation (modified julian day), telescope  
      projected separation, baseline projection angle  
      and squared visibility}  
\end{deluxetable}  
  
\clearpage  

\begin{deluxetable}{ccc}  
  \tablecaption{$\alpha$~Per PHOENIX models.\label{phoenix_models}}  
  \tablehead{  \colhead{$T_\mathrm{eff}$} & \colhead{log g = 1.4} 
    & \colhead{log g = 1.7} \\ (K)}    
  \startdata  
  6150& 0.137       &   0.136     \\  
  6270& 0.135       &   0.134     \\  
  6390& 0.133       &   0.132       
  \enddata  
  \tablecomments{Models tabulated for  
    different effective temperatures and surface gravities. The  
    $K$-band CLD is condensed into a power law coefficient $\alpha$:  
    $I(\mu)\propto \mu^\alpha$.}  
\end{deluxetable}

\clearpage  

\begin{deluxetable}{llllc}  
  \tablecaption{Models for $\alpha$~Per. \label{mirfak_models}}  
   \tablehead{  
     \colhead{Model} & \colhead{$\theta_\star$} &  
     \colhead{$\alpha$} & \colhead{$F_s/F_\star$} &  
     \colhead{$\chi^2_r$} \\ & (mas) & & (\%)  }  
   \startdata  
   UD             & $3.080_{\pm 0.004}$ & 0.000   & -  & 5.9 \\   
   PHOENIX CLD    & $3.137_{\pm 0.004}$ & 0.135   & -  & 1.1\\   
   adjusted CLD   & $3.130_{\pm 0.007}$ & $0.119_{\pm 0.016}$ & - & 1.0\\   
   \hline  
   CSE ``Polaris''& $3.086_{\pm 0.007}$ & $0.135$ & $1.5$ & 3.0 \\  
   CSE (1)&$3.048_{\pm 0.007}$ & $0.066_{\pm0.004}$ & $1.5$ & 1.4 \\  
   CSE (2)&$3.095_{\pm 0.010}$ & $0.135$ & $0.06_{\pm0.26}$ & 1.4   
   \enddata  
   \tablecomments{The parameters are: $\theta_\star$ the stellar  
     angular diameter, the CLD power law coefficient $\alpha$ and, if  
     relevant, the brightness ratio between the CSE and the star  
     $F_s/F_\star$. The first line is the uniform disk diameter, the  
     second one expected CLD from the PHOENIX model, the third one is  
     the adjusted CLD. The fourth line is a scaled Polaris CSE model  
     (Paper II). The last two lines are attempts to force a CSE model  
     to the data. Lower script are uncertainties, the absence of lower  
     script means that the parameter is fixed}  
\end{deluxetable}  

\clearpage  
  
\begin{deluxetable}{lcc}  
  \tablecaption{Relative flux in $K$-Band for the CSE discovered around  
    Cepheids and the non pulsating $\alpha$~Per.\label{cepheids_CSE}}  
  \tablehead{ \colhead{Star} & \colhead{Period} &  \colhead{CSE flux} \\
  & (d) & (\%)}  
  \startdata  
  $\alpha$~UMi &  3.97  &  $1.5\pm0.4$  \\
  $\delta$~Cep &  5.37  &  $1.5\pm0.4$  \\  
  Y~Oph        & 17.13  &  $5.0\pm2.0$  \\  
  $\ell$~Car   & 35.55  &  $4.2\pm0.2$  \\  
  \hline  
  $\alpha$~Per &     -  & $<0.26$        
  \enddata  
 \end{deluxetable}  

\end{document}